\documentclass[a4paper]{jpconf}
\usepackage{graphicx}

\usepackage{cite}
\usepackage{amssymb,amsmath}

\input epsf.sty
\newcommand{\BEQ}{\begin{equation}}          % Gleichungen Anfang ..
\newcommand{\BEA}{\begin{eqnarray}}
\newcommand{\EEQ}{\end{equation}}            % .. und Ende
\newcommand{\EEA}{\end{eqnarray}}
\newcommand{\D}{{\rm d}}                     % gerades d fuer Ableitungen
\newcommand{\II}{{\rm i}}                    % gerades i fuer komplexe Einheit
\newcommand{\wit}[1]{\widetilde{#1}}         % weite Schlange
\renewcommand{\vec}[1]{\boldsymbol{#1}}      % Vektoren fettgedruckt

\catcode`\@=11
\def\numberbysection{\@addtoreset{equation}{section}
                     \def\theequation{\thesection.\arabic{equation}}}
\def\up#1{\raise 1ex\hbox{\sevenrm#1}}
\def\build#1_#2^#3{\mathrel{\mathop{\kern 0pt#1}\limits_{#2}^{#3}}}

\numberbysection

\newpage
\begin{document}
\title{Lie symmetries of semi-linear Schr\"odinger equations and applications}

\author{Stoimen Stoimenov$^{a,b}$ {\rm and} Malte Henkel$^{a}$}

\address{$^a$Laboratoire de Physique des Mat\'eriaux (CNRS UMR 7556), Universit\'e Henri Poincar\'e Nancy I,
B.P.239, F-54506 Vand{\oe}uvre l\`es Nancy Cedex, France}

\address{$^b$Institute of Nuclear Research and Nuclear Energy,
Bulgarian Academy of Sciences,1784 Sofia, Bulgaria}

%\ead{stoimenov@lpm.u-nancy.fr}

\begin{abstract}
Conditional Lie symmetries of semi-linear $1D$ Schr\"odinger \index{Schr\"odinger equation}
\index{Schr\"odinger algebra} \index{parabolic subalgebra} and diffusion 
equations are studied if the mass (or the diffusion constant) is considered 
as an additional variable. In this way, dynamical symmetries of
semi-linear Schr\"odinger equations become related to the parabolic
and almost-parabolic subalgebras of a three-dimensional conformal Lie 
\index{conformal algebra} \index{conditional symmetry}
\index{ageing} \index{phase-ordering kinetics} algebra
$(\mathfrak{conf}_3)_{\mathbb{C}}$. The corresponding representations
of the parabolic and almost-parabolic subalgebras of
$(\mathfrak{conf}_3)_{\mathbb{C}}$ are classified and the
complete list of conditionally invariant 
semi-linear Schr\"odinger equations is obtained. 
Applications to the phase-ordering kinetics of simple magnets and to
simple particle-reaction models are briefly discussed.
  
\end{abstract}

\section{Introduction}

The understanding of the long-time kinetics of non-equilibrium continues to pose
challenging problems. Here we are interested in the long-standing problem of
phase-ordering kinetics of simple magnets which arises when, say, a ferromagnet is rapidly quenched from a disordered initial state to below its critical 
temperature $T<T_c$. We restrict attention here to the case with a 
non-conserved order-parameter, when a common description \cite{Bray94} is given in terms of a Langevin equation for the  
coarse-grained order-parameter $\Phi$ (referred to as `model A') 
\BEQ \label{eq:modelA}
\frac{\partial \Phi}{\partial t} = \Gamma \vec{\nabla}^2 \Phi 
-\frac{\D V(\Phi)}{\D \Phi} + \eta
\EEQ
where $V$ describes the self-interaction of $\Phi$, $\eta$ is 
a gaussian delta-correlated thermal noise $\eta$ of variance $T$ and 
white-noise initial conditions (which come from the disordered initial state) 
are assumed. The task is
to obtain the long-time behaviour of solutions of equation (\ref{eq:modelA}); in particular one 
would like to be able to understand the origin of the dynamical scaling 
which has been observed in numerous exactly solvable systems, in many numerical simulations and
in several experiments, see \cite{Bray94,Cugl02,Godr02,Henk04b} and references 
therein. These results are conveniently formulated in terms of the two-time
autocorrelation and the (linear) autoresponse functions defined as
\BEA
C(t,s) &:=&\langle \Phi(t) \Phi (s)\rangle \:=\: s^{-b}f_C({t/s}) \nonumber \\
R(t,s) &:=& \left.{\delta\langle \Phi (t)\rangle \over \delta h(s)}\right|_{h=0} \:=\: s^{-1-a}f_R({t/s})
\EEA
where $h(s)$ is time-dependent magnetic field conjugate to the non-conserved order-parameter $\Phi$. 
We have also included here the conventionally admitted scaling forms, which are valid in the double scaling
limit $t,s\gg t_{\rm micro}$ and $t-s\gg t_{\rm micro}$ where $t_{\rm micro}$ is some `microscopic' reference time. The {\em ageing behaviour} of the system 
is expressed through the
breaking of time-translation invariance, that is $C(t,s)$ or $R(t,s)$ depend on both $t$ and $s$
and {\em not} merely on the difference $t-s$. Physically, ageing comes from the existence of at
least two stable, equivalent and competing stationary states of the system, 
such that although locally  a stationary
state is rapidly reached, globally the system never manages to decide to which of the possible stationary
states it should relax \cite{Bray94,Cugl02,Godr02,Henk04b}. 

In view of the complexity of the problem, an approach based on the
dynamical symmetries of the ageing phenomenon would be helpful. Such an approach is 
possible since the behaviour of solutions of (\ref{eq:modelA}) actually only depends
on the deterministic part of that equation, provided that this deterministic 
part is Galilei-invariant.\\ 

\noindent 
{\bf Theorem.}\cite{Pico04} {\it Consider the deterministic part of eq.~(\ref{eq:modelA}), which is
obtained by setting $\eta=0$. If that deterministic part is Galilei-invariant, then all
$n$-point correlation and response functions of the full stochastic theory can be exactly
expressed in terms of $(n+1$)- or $(n+2)$-functions determined from the
deterministic part alone.}\\ 

In particular, the two-time response function $R(t,s)$ can in this way be shown to be independent of the
noises in eq.~(\ref{eq:modelA}). The requirement of Galilei-invariance in phase-ordering kinetics is
a natural one, because the dynamical exponent $z=2$ in this case \cite{Bray94}. Furthermore, the
form of $R(t,s)$ and also of $C(t,s)$ can be predicted is found in very good agreement with
simulational data \cite{Henk03c,Henk04,Lore05}. 
In consequence, for an understanding of the dynamical symmetries
of phase-ordering kinetics it is sufficient to study the symmetries of the deterministic reaction-diffusion
equation following from eq.~(\ref{eq:modelA}).
Here, we take up the question of how to establish non-trivial dynamical symmetries of such deterministic partial
differential equations.

\section{Dynamical symmetries with fixed masses}
As a first example of our procedure, we consider the case when the
`mass' $\cal M$ is a fixed constant. The non-linear Schr\"odinger equation 
reads (for simplicity in $d=1$ space dimension)
\BEQ\label{eq:NLSE}
2{\cal M} \partial_t\Phi-\partial^2_{\vec r}\Phi =gF(t,r,\Phi , \Phi^*)
\EEQ
where $F$ is referred to as a potential and $g$ is a coupling constant. If $F=0$, the
equation becomes linear and its well-known dynamical symmetry is given by the
Schr\"odinger algebra $\mathfrak{sch}_1$ \cite{Nied72}, spanned by the 
generators 
\BEA
X_{-1} &=& -\partial_t \;\; , \;\; Y_{-{1\over 2}}=-\partial_r, \;\; M_0 =-{\cal M} \nonumber \\
Y_{1\over 2} &=&-t\partial_r-{\cal M} r \; \;,\;\; X_0 =-t\partial_t-{1\over 2}r\partial_r-{x\over 2}
\label{eq:sch1} \\
X_1 &=& -t^2\partial_t-tr\partial_r-{{\cal M}\over 2}r^2-xt
\nonumber 
\EEA
It is also well-known that, for a dimensionless $g$, 
the non-linear equation is only Schr\"odinger-invariant for the special
choice of the potential $F=\left(\Phi\Phi^*\right)^{2}\Phi$ \cite{Fush93}. 
  
However, the equation (\ref{eq:NLSE}) has in general complex solutions
while for an application to the kinetic equation (\ref{eq:modelA}) one needs real-valued solutions. Furthermore, the available empirical evidence in
favour of Schr\"odinger-invariance in the kinetic Ising model in
$d=1,2$ and $3$ \cite{Henk03c,Pico04} goes against the mathematical result cited above. 
In order to overcome these difficulties, we recognize that $g$ is in general a dimensionful quantity and hence 
we shall look for the dynamical symmetries algebra of (\ref{eq:NLSE}) taking this into account \cite{Stoi05,Baum05}. 

If the dimensionful coupling constant enters into theory, the generators  (\ref{eq:sch1}) are in general modified
by additional $g$-dependent terms. We take the generator of scale-transformation 
in the form \cite{Baum05} 
\BEQ\label{eq:MSC}
X_0 =-t\partial_t-{1\over 2}r\partial_r-yg\partial_g-{x\over 2}, 
\EEQ
where the dimensionful coupling $g$ enters together with its scaling
dimension $y$, and other generators are calculated such that the
commutators of $\mathfrak{sch}_1$ are kept and the 
free Schr\"odinger operator ${\cal S}=2M_0 X_{-1} - Y_{-1/2}^2$ remains unchanged in this new representation. We find that only the generator of special transformations is modified with respect to (\ref{eq:sch1})
\BEQ \label{eq:X1sfinal}
X_1 = - t^2 \partial_t - t r\partial_t - 2 {y} t g\partial_g -\frac{\mathcal{M} r^2}{2} - x t
\EEQ
Furthermore if the time-translation invariance is broken, merely 
invariance under the {\it ageing algebra}
${\mathfrak{age}}_1 := \langle X_{0,1}, Y_{-\frac{1}{2},\frac{1}{2}}, M_0 \rangle $
should hold. In this case we obtain a more general form of special transformations
\BEQ \label{eq:X1agfin}
X_1 = - t^2 \partial_t - t r\partial_t - 2 {y} t g\partial_g -  m_0
g^{1+1/\hat{y}}\partial_g -\frac{\mathcal{M} r^2}{2} - x t,
\EEQ
where $m_0$ is constant which characterizes the representations of
$\mathfrak{age}_1 $ (for $\mathfrak{sch}_1$, $m_0=0$). 

Now standard methods \cite{Boye76} lead to the following form of non-linear part in equation (\ref{eq:NLSE}) \cite{Baum05}
\BEQ\label{eq:LIMIT}
F = \Phi \left( \Phi \Phi^*\right)^{1/x} f\left( \left( \Phi
\Phi^*\right)^{{y}}
\left[g^{-1/{y}}-\frac{m_0}{{y}\,t}\right]^{-x{y}}\right).
\EEQ
Consequently, in the long-time limit the equations invariant under
${\mathfrak{sch}}_1$ ($m_0=0$) and ${\mathfrak{age}}_1$ ($m_0\ne 0$) become
indistinguishable. It is a property of theories with fixed masses that
the potential still depends on {\em both} $\Phi$ and its `complex conjugate' $\Phi^*$. In applications to non-equilibrium dynamics, the latter
can be identified with the response field $\widetilde{\Phi}$ associated to the
order-parameter field. 
We have shown recently that the exact results obtained for the 
bosonic variants of the contact-process and the pair-contact process
for $C(t,s)$ and $R(t,s)$ \cite{Baum04,Baum05b} 
can be completely explained in terms of local scale-invariance, where
the action of the effective field-theory is split into a
Schr\"odinger-invariant term leading to an equation of motion with a non-linearity of the form (\ref{eq:LIMIT}) and a pure noise term \cite{Baum05}. 

\section{Dynamic symmetries with variable masses}

\begin{figure}[t]
\centerline{\epsfxsize=1.50in\ \epsfbox{
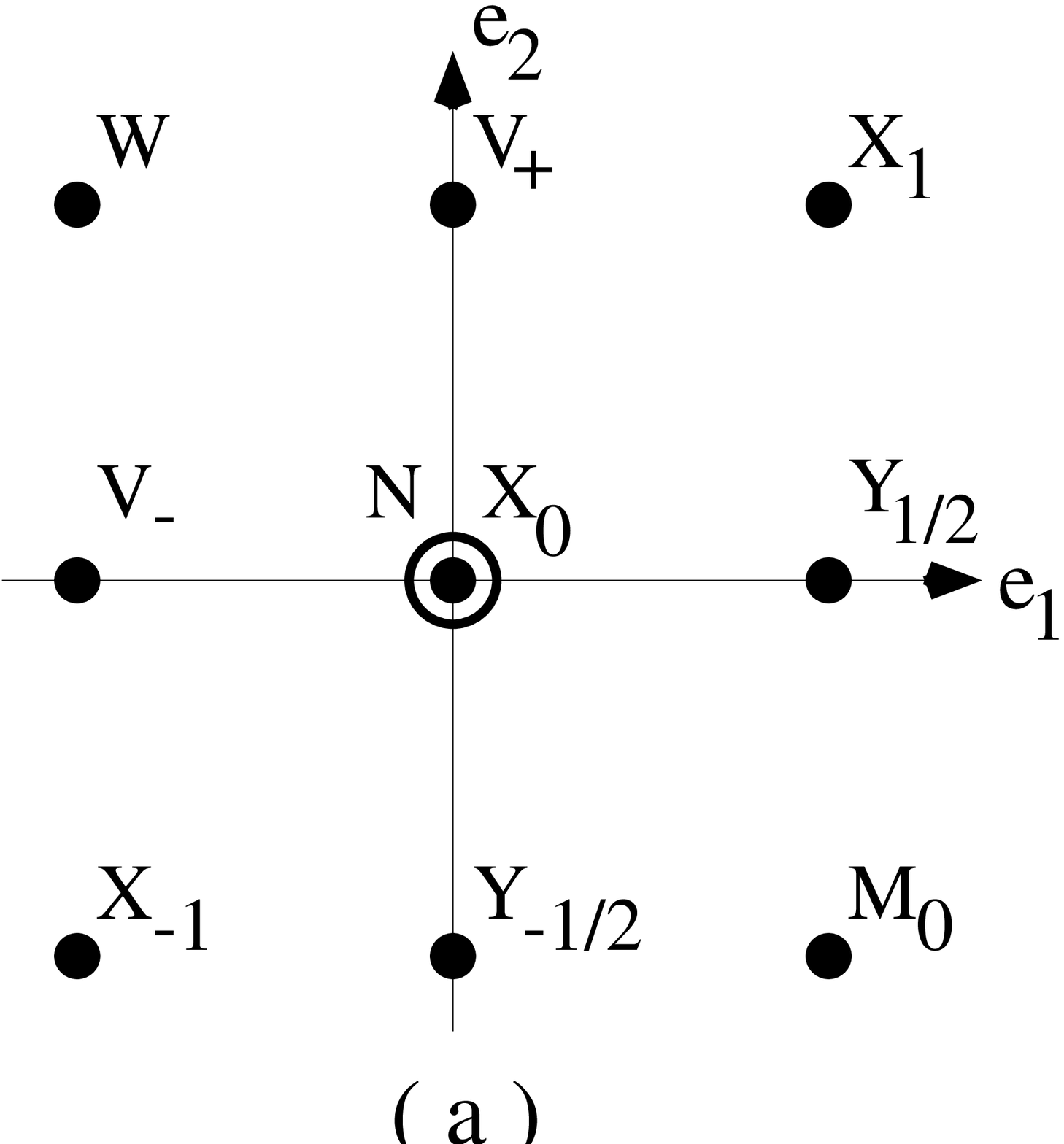} ~
\epsfxsize=1.50in\epsfbox{
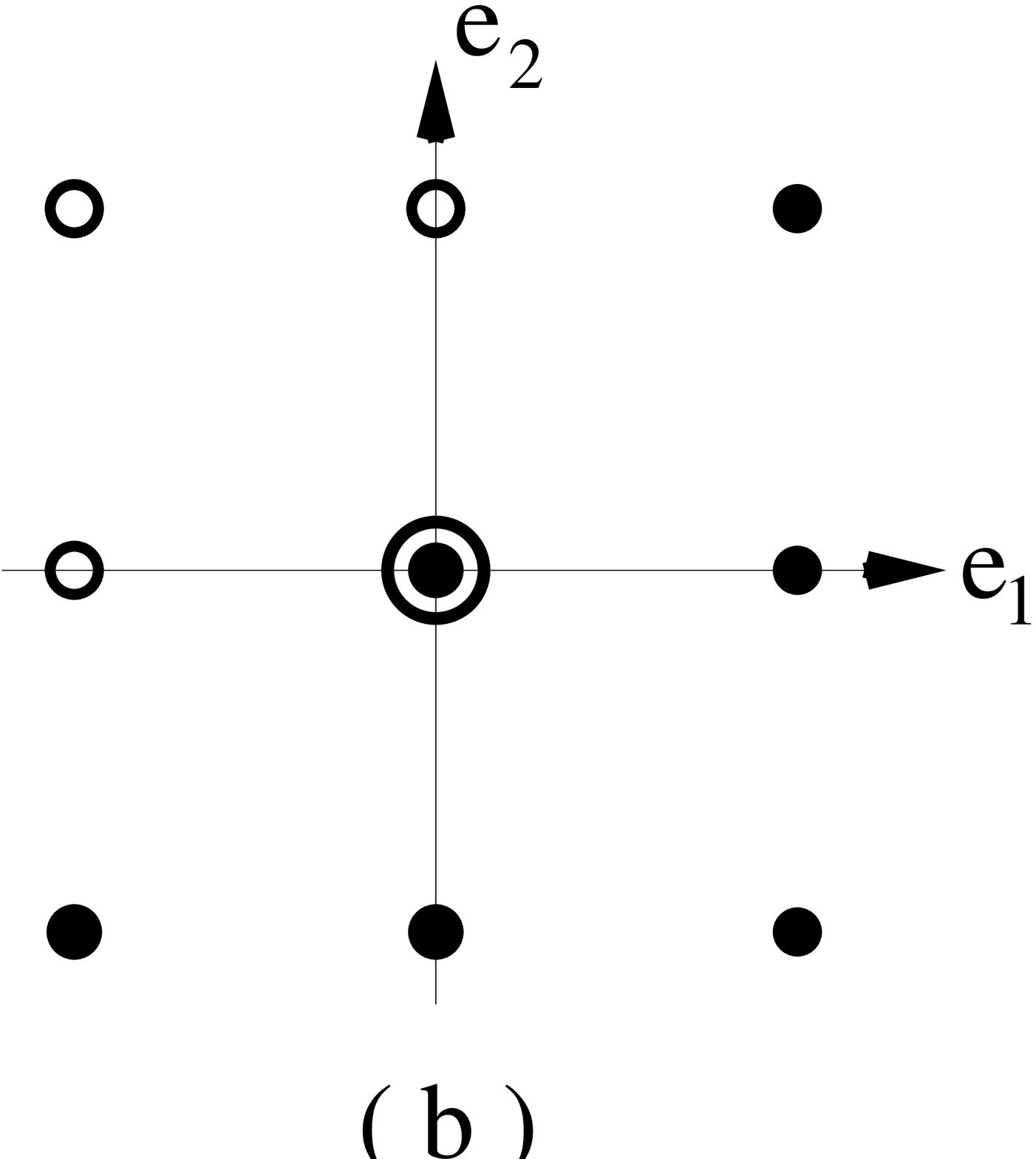} ~
\epsfxsize=1.50in\epsfbox{
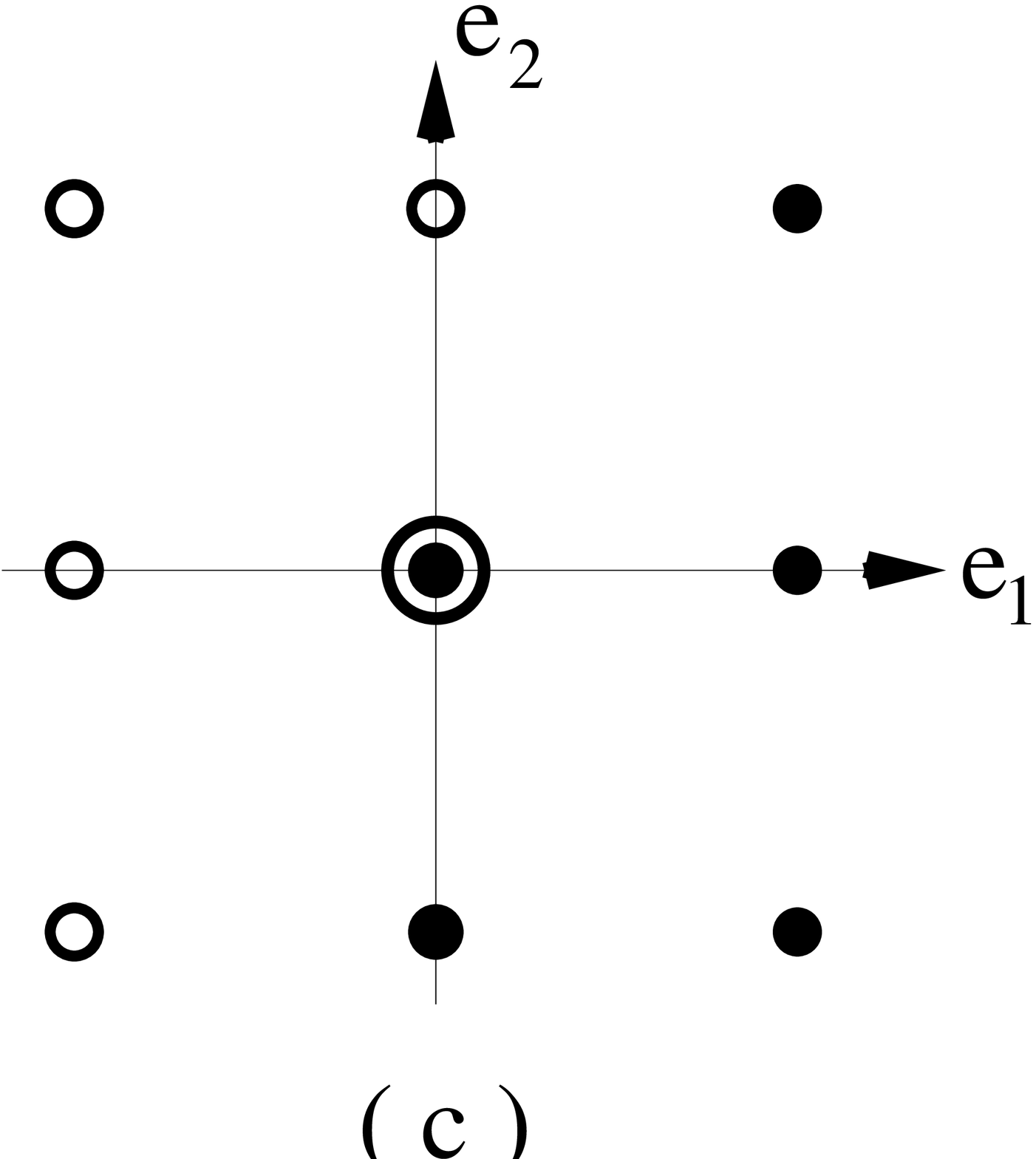} ~
\epsfxsize=1.50in\epsfbox{
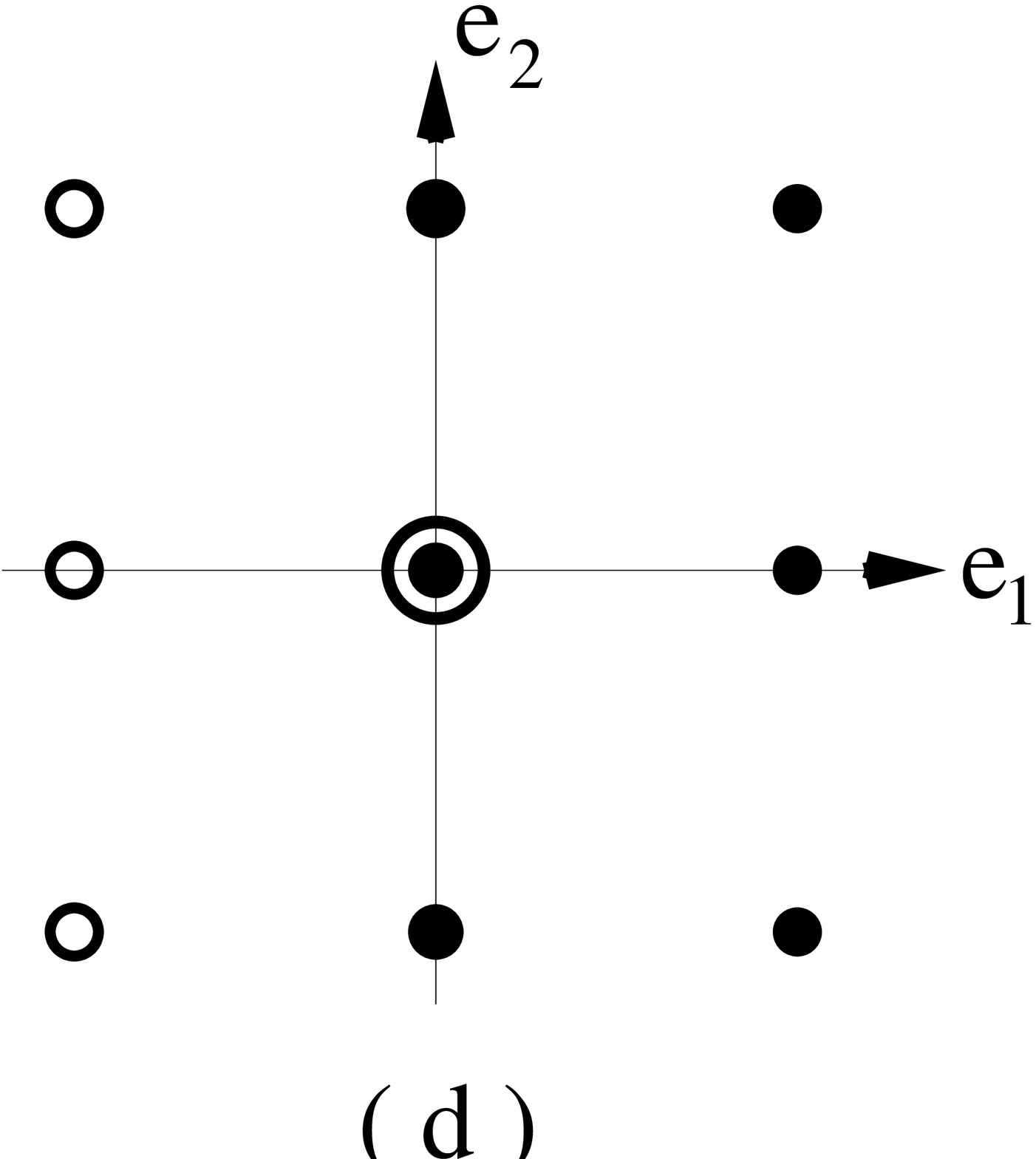}
}
\caption[Root space]{(a) Root diagram of the complex Lie algebra $B_2$ and the 
identification of the generators  
of the complexified conformal Lie algebra
$(\mathfrak{conf}_3)_{\mathbb{C}}\supset(\mathfrak{sch}_1)_{\mathbb{C}}$.
The double circle in the center denotes the Cartan subalgebra.   
The generators of the three non-isomorphic parabolic subalgebras 
are indicated by the full points, namely
(b) $\wit{\mathfrak{sch}}_1$, (c) $\wit{\mathfrak{age}}_1$ and
(d) $\wit{\mathfrak{alt}}_1$. 
\label{Bild1}}
\end{figure}

We now extend the scope of our investigation and also consider the
`mass' $\cal M$ as a new coordinate in the problem. 
Then the Schr\"odinger algebra 
$\mathfrak{sch}_d$ in $d$ spatial dimensions can be embedded into 
the complexified conformal algebra
$(\mathfrak{conf}_{d+2})_{\mathbb{C}}$ in $d+2$ dimensions. One may 
see this in a simple way by writing  ${\cal M}=\II m$ and then performing 
a Fourier transformation which defines a new wave function \cite{Henk03b}
\BEQ \label{eq:Fourier}
\Phi= \Phi_m(t,\vec{r})=
{1\over \sqrt{2\pi\,}}\int_{\mathbb{R}}\!\D\zeta\:
e^{-\II m\zeta}\Psi(\zeta,t,\vec{r})
\EEQ
The one-dimensional free Schr\"odinger equation then becomes
$\left(2\partial_{\zeta}\partial_t -\partial_r^2\right)\Psi=0$ which can be rewritten through a further change of variables as a massless Klein-Gordon/Laplace 
equation in three dimensions,\footnote{We point out that the `mass'
$\cal M$ is a {\em non-relativistic} mass, which plays a completely different 
r\^ole than the masses of relativistic theories.} which has the simple Lie algebra 
$(\mathfrak{conf}_3)_{\mathbb{C}}$ as a dynamical symmetry. 
The root diagram \index{root diagram} of the latter is shown in part (a) of figure~\ref{Bild1},
from which the correspondence between the
roots and the generators of $\mathfrak{sch}_1$ can be read off. 
Four additional generators $N, V_-, W, V_+$ should be added in order to
get the full conformal algebra $(\mathfrak{conf}_3)_{\mathbb{C}}$.

%%----------------------------------------------------------------------------%%
%----------------------------------------------------------------------------%%

In what follows, the so-called  
`parabolic subalgebras' of $\mathfrak{conf}_3$ are of interest. Up to 
isomorphisms, these are \cite{Henk03b}
\begin{itemize}
\item $\wit{\mathfrak{sch}}_1 := \langle X_{-1,0,1}, Y_{-\frac{1}{2},\frac{1}{2}}, M_0, N\rangle $,
see black dots in part (b) of the figure. 
\item $\wit{\mathfrak{age}}_1 := \langle X_{0,1}, Y_{-\frac{1}{2},\frac{1}{2}}, M_0, N\rangle$, 
see black dots in part (c) of the figure.
\item $\wit{\mathfrak{alt}}_1 := \langle D, X_1, Y_{-\frac{1}{2},\frac{1}{2}}, M_0, N, V_+\rangle$, 
see part (d) of the figure.
\end{itemize}
Here the generator $D=2X_0-N$ of the full dilatations is used. 
The corresponding {\it ``almost-parabolic subalgebras''} \cite{Stoi05} are
the same as above but without the generator $N$.

For notational simplicity we shall work in one space dimension and look for a semi-linear extension of the
linear  ``Schr\"odinger equation'' of the form \cite{Stoi05}
\BEQ \label{gl:Sop}
\hat{\cal S}\Psi := \left( 2\partial_{\zeta}\partial_t-\partial^2_r\right)\Psi 
= F(g,\zeta,t,r,\Psi ,\Psi^*)
\EEQ
which is invariant under one of the parabolic subalgebras of 
$(\mathfrak{conf}_3)_{\mathbb{C}}$. We first 
construct new differential-operator representations of the
algebras defined above which contain also a dimensionful coupling $g$.
Then we explicitly give the admissible forms of $F$ in (\ref{eq:NLSE}).

\subsection{Invariant linear equations}
The new representations with dimensionful coupling
constants of (almost) parabolic subalgebras
of $(\mathfrak{conf}_3)_{\mathbb{C}}$ are constructed as follows. As before, 
we take for the dilatation generators
\BEQ\label{eq:SC}
X_0 =-t\partial_t-{1\over 2}r\partial_r-yg\partial_g-{x\over 2}\;\;,\;\;D =-t\partial_t-r\partial_r-\zeta \partial_{\zeta}-sg\partial_g-x 
\EEQ
The exponents $y$ and $s$ describe the scaling behaviour of the coupling $g$. Space- and time-translations  $Y_{-{1\over 2}}=-\partial_r, X_{-1}=-\partial_t$ are also fixed, while the remaining generators are taken in the form
\BEA
M_0 &=&-\partial_{\zeta}-L(t,r,\zeta ,g)\partial_g \;\;,\;\;Y_{1/2} = -t\partial_r-r\partial_{\zeta }
                 -Q(t,r,\zeta ,g)\partial_g \nonumber \\
X_1 &=& -t^2\partial_t-tr\partial_r-{1\over 2}r^2\partial_{\zeta}
                 -P(t,r,\zeta ,g)\partial_g-xt \nonumber \\
N &=& -t\partial_t+\zeta \partial_{\zeta }
      -K(t,r,\zeta ,g)\partial_g 
\\
V_+ &=& -2tr\partial_t-2\zeta r\partial_{\zeta }-(r^2+2\zeta t)\partial_r
        -F(t,r,\zeta ,g)\partial_g-2xr. \nonumber 
\EEA
The unknown functions $L,Q,P,K,F$ are found from the commutation relations
and the requirement that the linear equation $\hat{S}\Psi=0$ is left invariant under above transformations.
Our results are presented in the following table.
%%++++++++++++++++++++++++++++++++++++++++++++++++++++++++++++++++++++++++++++++
\begin{tabular}{|||l|l||l|l|l|||} \hline\hline\hline
case & $\mathfrak{g}$ & representation & $x$ & $\hat{S}$ \\ \hline\hline\hline
%0 & $\mathfrak{age}_1$ & NMG & $=1/2$ & $2\partial_{\zeta}\partial_t-\partial_r^2$ \\
 %& & $L=0$, $Q=0$, & &    \\
  %& & $P=p_{01}m_0 g^{(y+1)/y}$ &  &   \\ \cline{3-5}
  %& & NMG & $\ne 1/2$ & $2\partial_{\zeta}\partial_t-\partial_r^2$  \\
  %& & $L=0$, $Q=0$,   &  & \\
  %& & $P=p_{01}t^{y+1}m(t^y/g)$ & & $\partial_r^2$ \\ \hline 
1 & $\mathfrak{age}_1$ & NMG & $=1/2$ & $2\partial_{\zeta}\partial_t-\partial_r^2$ \\
\cline{4-5} 
  & & $L=0$, $Q=0$, & $\ne 1/2$ & $2\partial_{\zeta}\partial_t-\partial_r^2$ \\
  & & $P=p_{01}tg$ & & $\partial_r^2$ \\
\cline{1-1}\cline{3-3} 
2 & $\wit{\mathfrak{age}}_1$ & $K=k_0 g$ &  & \\ \hline \hline
3 & $\mathfrak{age}_1$ & MMG & $=1/2$ & $2\partial_{\zeta}\partial_t-4yg\zeta^{-1}\partial_g\partial_{t}-\partial_r^2$ \\ \cline{4-5}
  & & $L=-2y\,g/\zeta$  & &  \\
  & & $Q=-2y\,gr/\zeta$ & $\ne 1/2$ & $(2\partial_{\zeta}\partial_t-\partial_r^2)\Psi=0$ \\
  & & $P=-ygr^2/\zeta$ &  & \\ \cline{1-1}\cline{3-3} 
4 & $\wit{\mathfrak{age}}_1$ & $K=k_0 g$ & & \\ \hline\hline 
5 & $\mathfrak{alt}_1$ & $L=sg/\zeta$, $Q=srg/\zeta$ & $=1/2$ & $2\partial_\zeta\partial_t+2sg\zeta^{-1}\partial_g\partial_t-\partial_r^2$ \\
  & & $P=sr^2g/2\zeta$ &  & \\ \cline{4-5}
  & & $F=2srg$ & $\ne 1/2$ &
      $\partial_{\zeta}\partial_t$ \\ \cline{1-1}\cline{3-3}
6 & $\wit{\mathfrak{alt}}_1$ & $K=k_{0}' g$ & & \\ \hline\hline
7 & $\mathfrak{sch}_1$ & $L=Q=P=0$ & $=1/2$ & $2\partial_{\zeta}\partial_t-\partial_r^2$ \\ 
  & & & & $\partial_r^2$ \\ \cline{3-5}
  & & $L=Q=0$, $P=2ytg$ & $\ne 1/2$ & $\partial_r^2$ \\ 
  & & & & $2\partial_{\zeta}\partial_t-\partial_r^2$ \\ 
\cline{1-1}\cline{3-3}
8 & $\wit{\mathfrak{sch}}_1$ & $K=k_0g$ & & \\ \hline\hline\hline
\end{tabular} \\
%%++++++++++++++++++++++++++++++++++++++++++++++++++++++++++++++++++++++++++++++

\noindent 
We obtain two distinct representations (see \cite{Stoi05} for details): \\ 
1. representations with $L=0$ for $\mathfrak{age}_1$, $\mathfrak{sch}_1$ which we call `non-modified' (NMG).\\
2. representations with $L\ne 0$ for $\mathfrak{age}_1$, $\mathfrak{alt}_1$ which we call `modified' (MMG). \\
We see that for $x\ne 1/2$ auxiliary condition(s) lead to modified forms of $\hat{S}$ \cite{Stoi05}. 

\subsection{Invariant non-linear equations}
{}From these representations, standard methods \cite{Boye76,Fush93} allow to calculate 
the potential $F$ in the non-linear invariant equation $\hat{S}\Psi =F$. 
Our results \cite{Stoi05} are presented in the following table where we give 
\begin{enumerate}
\item Generic solutions for the potential $F$. 
\item Non-generic solutions with certain conditions on parameters 
of the algebras and where $f$ stands for an arbitrary function.
\end{enumerate}

%%++++++++++++++++++++++++++++++++++++++++++++++++++++++++++++++++++++++++++++++
\hspace{-0.8truecm}
\begin{tabular}{||l|l||l|l||} \hline \hline
case & subalgebra  & potential $F$ & condition  \\ \hline \hline
%0 & $\mathfrak{age}_1$ & $\Psi^{x+\frac{2}{x}}f\left(\ln\Psi +\int{\!\D v\over v}{p_{01}vm(v)-2y\over %p_{01}vm(v)-y},\Psi/\Psi^*\right)$ \\ \hline
1 & $\mathfrak{age}_1$ & $a^{x+2}f(a^x\Psi, \Psi/\Psi^*)$ & \\
2 & $\wit{\mathfrak{age}}_1$ & $\Psi^{(x+2)/x}f(\Psi/\Psi^*)$ & $p_{01}\ne2y-k_0$  \\ 
  & $\wit{\mathfrak{age}}_1$  & $a^{x+2}f(a^x\Psi, \Psi/\Psi^*)$ &
    $p_{01}=2y-k_0$ \\ \hline
3 & $\mathfrak{age}_1$ & $b^{(x+2)}f(b^{-x}\Psi,\Psi/\Psi^*)$ &  \\
4 & $\wit{\mathfrak{age}}_1$ & $\Psi^{(x+2)/x}f(\Psi/\Psi^*)$ & $k_0\ne 4y$ \\ 
  & $\wit{\mathfrak{age}}_1$ & $b^{(x+2)}f(b^{-x}\Psi,\Psi/\Psi^*)$ & $k_0=4y$ \\ \hline
5 & $\mathfrak{alt}_1$ & $t^{-x-2}f(\zeta^{-s}g,t^x\Psi,\Psi/\Psi^*)$ & \\
6 & $\wit{\mathfrak{alt}}_1$ & $c^{-x-2}f(c^x\Psi,\Psi/\Psi^*)$ & \\ \hline
7 & $\mathfrak{sch}_1$ & $g^{-(x+2)/2y}f(g^{x/2y}\Psi,\Psi/\Psi^*)$ &  \\
8 & $\wit{\mathfrak{sch}}_1$ & $\Psi^{(x+2)/x}f(\Psi/\Psi^*)$ & $k_0\ne 0$  \\ 
  & $\wit{\mathfrak{sch}}_1$ & $g^{-(x+2)/2y}f(g^{x/2y}\Psi,\Psi/\Psi^*)$ & $k_0=0$  \\ \hline\hline
\end{tabular}

\section{An application to phase-ordering}

Consider the equations invariant under $\mathfrak{age}_1$ or 
$\mathfrak{sch}_1$ with
a corresponding auxiliary condition for the wave function. 
If the potential does not depend on the second variable $\Psi/\Psi^*$ (a 
`phase') we obtain the following invariant semi-linear equation \cite{Stoi05}
\BEQ
\left( 2\partial_{\zeta}\partial_t - \partial_r^2\right) \psi = \psi^5 \bar{f}\left(g \psi^{4y}\right)
\EEQ
where $\bar{f}$ is an arbitrary function (see cases 2, 8). 
This kind of invariant equations with a {\em real} potential may be of interest 
in phase-ordering kinetics, see \cite{Bray94,Cugl02,Godr02}. We point out that
the representation of the invariance algebra 
we found is the same for {\em any} choice of the function 
$\bar{f}$, which is in keeping with the expected universality. 

Summarizing, taking into account that the couplings which arise in 
semi-linear equation are in general dimensionful has allowed us to
extend previous results on the existence of non-trivial dynamical symmetries
of these equations, which bring the mathematical theory into much closer
contact with simulational results. In particular, going over to variable
masses appears to bring the possibility of actually proving dynamical scaling
and also further non-anticipated symmetries within reach. We hope to return
to this in the future.

We thank F. Baumann, R. Cherniha and R. Schott for useful conversations. 
S.S. was supported by the EU Research Training Network HPRN-CT-2002-00279.

\end{document}